\documentstyle[12pt]{report} 
\newcommand{\be}{\begin{equation}} 
\newcommand{\ee}{\end{equation}} 
\newcommand{\bea}{\begin{eqnarray}} 
\newcommand{\eea}{\end{eqnarray}} 
\newcommand{\sss}{\scriptscriptstyle} 
\renewcommand{\S}{{\cal S}} 
 
\newcommand{\Bthree}{\mbox{${}^3\! B$}} 
\newcommand{\Nbar}{{\bar N}} 
\newcommand{\ubar}{{\bar u}} 
\newcommand{\Kbar}{{\bar K}} 
\newcommand{\alphabar}{{\bar\alpha}} 
\newcommand{\Vbar}{{\bar V}} 
\newcommand{\Pbar}{{\bar P}} 
\begin{document} 
%%%%%%%%%%%%%%%%%%%%%%%%%%%%%%%%%%%%%%
\begin{titlepage} 
\begin{flushright} 
  IFP--UNC--491\\  
  TAR--UNC--043\\ 
  CTMP/007/NCSU\\ 
\end{flushright}
\vspace{0.6in}
\begin{center} {\large The Path Integral Formulation of Gravitational 
Thermodynamics\footnote{Based on the talk presented by J.D. Brown at the 
conference {\it The Black Hole 25 Years After}, Santiago, Chile, January
1994.}} \\
\vspace{.5in} 
J. David Brown\\
Departments of Physics and Mathematics\\
North Carolina State University\\ 
Raleigh, NC 27695--8202\\
\vspace{.2in} 
James W. York\\ 
Department of Physics and Astronomy\\ 
The University of North Carolina\\ 
Chapel Hill, NC 27599--3255\\
\vspace{.5in} 
%%%%%%%%%%%%%%%%%%%%%%%%%%%%%%%%%%%%%
%Abstract
\end{center} 
%
%\noindent This is the abstract. This is the abstract. This is the abstract. 
%This is the abstract. This is the abstract. This is the abstract.
\end{titlepage} 
%%%%%%%%%%%%%%%%%%%%%%%%%%%%%%%%%%%%%%%%%%%%%%%%%%%%%%%%%%%%%%%%%%%%%%%%
\addtocounter{chapter}{1}
\centerline{\bf I. Introduction} 
\vspace{4pt} 
In the early 1970's Bekenstein argued that black holes have entropy [1] 
and Hawking showed that black holes have temperature [2]. A few years 
later these conclusions were apparently 
confirmed by Gibbons and Hawking [3] who used path integral methods to
evaluate the black hole partition function. However, the original 
analysis of Gibbons and Hawking is incorrect. The
difficulty arises because, in the language of ordinary thermodynamics, 
black holes are unstable and 
have a negative heat capacity. On the other hand, any system that can be
described by a 
partiton function is necessarily stable and has a positive heat capacity. 

The path integral approach of Gibbons and Hawking is also somewhat mysterious
regarding the origin of black hole entropy. A number of authors have 
speculated that the large 
entropy of a black hole should be associated with a large number of internal
states, hidden by the horizon, that are consistent with the few external 
parameters that characterize 
the black hole. Others have argued that black hole entropy can be associated
with a large number of possible initial states that can collapse to form a 
given black hole. 
The path integral analysis does not support either of these views in an
obvious way. The key 
element in the path integral calculation is the action of a static Euclidean
black hole geometry. A static Euclidean black hole has no horizon, no 
interior region, and no 
information about the collapse process. The path integral calculation of the
partition function seems to hint at a topological explanation for black hole
entropy. 

The first objective of this article is to show that the black hole partition
function 
can be placed on a firm logical foundation by enclosing the black hole in a
spatially 
finite ``box" or boundary. The presence of the box has the effect of
stabilizing the 
black hole and yields a system with a positive heat capacity. Without a 
finite box, or some other mechanism for stabilizing the black hole (such 
as a negative cosmological constant [4]), the partition function does not 
exist. It is, nevertheless, common practice to fix boundary conditions at 
infinity and treat the zero--loop approximation to the path integral as if 
it were a real partition function. Such an identification is logically 
unfounded and, at any rate, cannot be extended beyond the zero--loop 
approximation. 

The second objective 
of this article is to explore the origin of black hole entropy. This is 
accomplished through the construction of a path integral expression for 
the density matrix for 
the gravitational field, and through an analysis of the connection between 
the density 
matrix and the black hole density of states. (The density matrix for a black 
hole has been studied previously in Ref.~[5].) Our results suggest that 
black hole entropy can be associated with an absence of certain ``inner 
boundary information" for the system. 

We begin in Sec.~2 with a review of the motivations for enclosing the black 
hole in a box. In Sec.~3 we analyze the action for the gravitational field 
in the presence of a spatially finite boundary. 
Section 4 contains a discussion of the relationship between Lorentzian and 
Euclidean notation for the gravitational action. Formal path integral
expressions 
for the density matrix, density of states, and partition function are
constructed 
in Sec.~5. In Sec.~6 we use these results to calculate the density of states
for a black hole and we discuss the origin of black hole entropy. 
%%%%%%%%%%%%%%%%%%%%%%%%%%%%%%%%%%%%%%%%%%%%%%%%%%%%%%%%%%%%%%%%%%%% 

\bigskip
\addtocounter{chapter}{1}
\setcounter{equation}{0}
\vspace{10pt}
\centerline{\bf II. Finite Boundaries} 
\vspace{4pt}
Hawking's analysis [2] shows that the 
temperature of a black hole, as measured at spatial infinity, equals the
surface 
gravity divided by $2\pi$. For a Schwarzschild black hole of mass $M$, it
follows 
that the inverse temperature $\beta$ at infinity is $8\pi M$. On the other
hand, 
the standard thermodynamical definition of inverse temperature [6] is 
$\beta = \partial \S(E)/\partial E$, where $\S(E)$ is the entropy function
and $E$ is 
the thermodynamical internal energy. If the mass at infinity $M$ and the
internal 
energy $E$ are identified, then the relationship $\partial \S(E)/\partial 
E = 8\pi M$ can be integrated to yield $\S(E) = 4\pi E^2$ (plus an additive 
constant). This result is in complete agreement with 
the prediction made by Bekenstein [1] that a black hole has entropy
proportional 
to the area of its event horizon. 

The black hole entropy $\S(E) = 4\pi E^2$ is a convex function of $E$. This 
is characteristic of an unstable thermodynamical system [6]. In this case, 
the instability arises because energy and temperature are inversely related 
for black holes. Thus, if fluctuations cause a black hole to absorb an extra 
amount of thermal radiation from its environment, its mass will 
increase and its temperature will decrease. The tendency then is for the
cooler 
black hole to absorb even more radiation from its hotter environment, 
causing the black hole to grow without bound. 

These results can be reformulated within the context of statistical
mechanics.  
First consider the canonical partition function $Z(\beta)$ for an arbitrary
system. 
In general $Z(\beta)$ is a sum over quantum states weighted by the Boltzmann
factor 
$e^{-\beta E}$. If $\nu(E)$ is the density of quantum states with energy $E$,
then  
\be 
   Z(\beta) = \int dE\, \nu(E) e^{-\beta E} \ .
\ee 
The partition function can also be expressed as 
\be 
   Z(\beta) = \int dE e^{-I(E)}  \ , 
\ee 
where the ``action" is defined by $I(E) \equiv \beta E - \S(E)$ and the 
entropy function $\S(E)$ is the logarithm of the density of states: 
$\S(E) \equiv \ln\nu(E)$. (Dimensionful constant factors can be included in 
the logarithm as necessary.) The integral over $E$ can be evaluated in a
steepest 
descents approximation by expanding the action $I(E)$ to quadratic order 
around the stationary points $E^*(\beta)$, which satisfy  
\be 
   0 = \left.\frac{\partial I}{\partial E}\right|_{E^*} = \beta - 
   \left.\frac{\partial \S}{\partial E}\right|_{E^*} \ . 
\ee 
The Gaussian integral associated with a stationary point $E^*$ will converge 
if the second derivative of the action at $E^*$ is positive: 
\be 
   \left.\frac{\partial^2 I}{\partial E^2}\right|_{E^*} = 
   - \left.\frac{\partial^2 \S}{\partial E^2}\right|_{E^*} > 0 \ . 
\ee 
This condition shows that the entropy $\S(E)$ should be a concave function 
at the extremum $E^*$ in order for the Gaussian integral to converge. 

A further significance of the condition (2.4) can be seen as follows. 
In the steepest descents approximation, the expectation value of energy 
is $\langle E\rangle \equiv -\partial\ln Z/\partial\beta \approx E^*$ and 
the heat capacity is 
$C\equiv \partial\langle E\rangle/\partial \beta^{-1} 
\approx {\partial E^*}/{\partial \beta^{-1}} $. By differentiating Eq.~(2.3) 
with respect to $\beta$, we find  
\be 
    1 = \frac{\partial E^*}{\partial \beta} 
    \left.\frac{\partial^2\S}{\partial E^2}\right|_{E^*} \ . 
\ee 
Therefore the heat capacity is given by 
\be 
   C \approx -\beta^2 \left( \left.\frac{\partial^2\S}{\partial
E^2}\right|_{E^*} \right)^{-1} 
   = \beta^2 \left( \left.\frac{\partial^2 I}{\partial E^2}\right|_{E^*}
\right)^{-1} \ . 
\ee 
Thus, we see that in the steepest descents approximation the convergence of
the 
integral for the canonical partition function is equivalent to the
thermodynamical  
stability of the system (the concavity of the entropy), which in turn is
equivalent 
to the positivity of the heat capacity. 

For the black hole in particular, the entropy $\S(E) = 4\pi E^2$ is not a
concave 
function of the internal energy and the integral for the partition function
diverges.  
This can be seen in the path integral formalism as well. In that case, the
partition 
function is expressed as a functional integral over Euclidean 
metrics with periodic time, where the time period equals  
the inverse temperature at infinity [3]. The 
action is extremized by a Euclidean black hole metric, but the integration
about 
that stationary point diverges. Formally, such a divergent Gaussian integral 
yields an imaginary result. The Euclidean black hole extremum is properly 
interpreted as an instanton that dominates the semiclassical evaluation of
the 
rate of black hole nucleation from flat space at finite temperature [7]. 

The preceeding discussion shows that the canonical partition function
$Z(\beta)$ 
characterizes the thermal properties of thermodynamically {\it stable\/}
systems. 
For unstable systems $Z(\beta)$ can give information concerning the rate of
decay 
from a quasi--stable configuration (such as ``hot flat space" in the black
hole 
example), but it cannot be used to define thermodynamical properties such as 
expectation values, fluctuations, response functions, {\it etc\/}. Thus,
before the 
partition function can be used as a probe of black hole thermodynamics, 
it is first necessary to stabilize the black hole. It was recognized in
Ref.~[8] 
that a black hole is rendered thermodynamically stable by enclosing it in a 
spatially finite ``box" or boundary whose walls are maintained at a finite 
temperature. In this case the energy and the temperature at the boundary are 
not inversely related because of the blueshift effect of temperature in a
stationary gravitational field [9]. If the black hole absorbs an extra 
amount of thermal radiation from its environment, its energy will increase 
thereby increasing the 
gravitational blueshift. Although the temperature of the black hole as
measured 
at infinity is decreased by such a fluctuation, the temperature 
as measured at the boundary can increase due to the enhanced blueshift
effect. 
The hotter black hole will then give up its excess energy and return to its
stable equilibrium configuration. 

The stabilizing effect of a finite box can be confirmed by the following 
simple analysis. Consider a Schwarschild black hole of mass $M$ surrounded 
by a spherical boundary of radius $R$. The inverse temperature at infinity 
is $8\pi M$, so the inverse temperature at the boundary is blueshifted to 
$\beta = 8\pi M\sqrt{1-2M/R}$ [9]. On the other hand, the inverse 
temperature is defined by $\beta = \partial \S(E)/\partial E$, where again 
$\S(E)$ is the entropy as a function of internal energy $E$. Now, the 
entropy of the black hole, at least in the zero--loop (``classical") 
approximation, depends only on the black hole size and is unaffected by 
the presence, absence, or proximity [8] of a 
finite box. Thus, we have $\S(E) = 4\pi M^2$ as before.  By equating the 
two expressions for inverse temperature we find 
\be 
  8\pi M\sqrt{1-2M/R} = \frac{\partial (4\pi M^2)}{\partial E} \ . 
\ee 
In this case the energy $E$ and the mass $M$ as measured at infinity 
do not coincide. Equation (2.7) can be integrated to yield [8,10] 
\be 
   E = R - R\sqrt{1 - 2M/R} \ ,
\ee 
where, for convenience, the integration constant has been chosen so that
$E\to M$ 
in the limit $R\to\infty$ with $M$ fixed. The significance of this 
expression can be seen by expanding $E$ in powers of $GM/R$ (where 
Newton's constant $G$ 
is set to unity), with the result $E = M + M^2/(2R) + \cdots$. This 
shows that the internal energy inside the box equals the energy at 
infinity $M$ 
{\it minus} the binding energy $-M^2/(2R)$ of a shell of mass $M$ and radius 
$R$. The binding energy $-M^2/(2R)$ is the energy associated with the
gravitational field outside the box [10]. Also observe that the internal 
energy takes values in the range $0\leq E\leq R$. 

By solving Eq.~(2.8) for $M$ as a function of $E$, we obtain the entropy
function 
\be 
   \S(E) = 4\pi E^2 \left( 1 - E/(2R) \right)^2 \ . 
\ee 
First, note that the derivative $\partial\S/\partial E$ is a concave 
function of $E$ that vanishes at the extreme values $E=0$ and $E=R$. 
It follows that  
$\partial\S/\partial E$ has a maximum $\beta_{cr}$.  For $\beta > 
\beta_{cr}$ the equation $\beta = \partial\S/\partial E$ has no solutions 
for $E$. On 
the other hand, for $\beta < \beta_{cr}$, there are two solutions $E_1$ and 
$E_2$. At the larger of these two solutions, say, $E_2$, the second
derivative 
$\partial^2\S/\partial E^2$ is negative and the stability criterion (2.4) is 
satisfied. At the smaller of these two solutions, $E_1$, the second
derivative 
$\partial^2\S/\partial E^2$ is positive and the stability criterion (2.4) 
is violated. These considerations indicate that for a small box at low
temperature 
($\beta > \beta_{cr}$), the equilibrium configuration consists of flat 
space. For a large box at high 
temperature ($\beta < \beta_{cr}$), the stable equilibrium configuration
consists 
of a large black hole with energy $E_2$. The unstable black hole with 
energy $E_1$ is an instanton that governs the nucleation of black holes from 
flat space. In the limit $R\to\infty$, the stable black hole configuration is
lost and only the instanton solution survives [8,4]. 

In the following sections, we will develop the formal functional integral 
expressions for the density of states, the density matrix, and the canonical 
partition function for the gravitational field with a spatially finite
boundary $B$. For the partition function in particular, the inverse 
temperature will be 
fixed as a boundary condition on $B$. The use of a finite boundary has the 
following very important consequence: Because gravitational fields cause
temperature 
to redshift and blueshift, one must allow for the temperature to be fixed to 
different values at different points on $B$. In other words, gravitating
systems 
in thermal equilibrium are not characterized by a single temperature but
instead 
by a temperature {\it field} on the {\it boundary} of the system [11,12]. 
Correspondingly, the partition function is actually a {\it functional\/}
$Z[\beta]$ 
of the inverse temperature field on $B$. For some problems, such as the  
Schwarzschild black hole in a spherical box discussed above, it is possible 
to choose the temperature to be a constant on 
$B$. In those cases $B$ coincides with an isothermal surface for the 
system. However, experience with the Kerr black hole shows that this must be 
viewed as a particular choice of boundary conditions, not the most general
choice. 
What happens in the Kerr case [11] is that the angular velocity of 
the black hole with respect to observers who are at rest in the stationary 
time slices enters as a ``chemical" potential conjugate to angular momentum.
It turns out that the constant temperature surfaces and the constant 
angular velocity surfaces do not coincide. Therefore it is necessary to 
allow for {\it some} thermodynamical data, either the temperature or the 
chemical potential or both, 
to vary across the boundary. This conclusion might seem disturbing at first, 
since traditionally one of the purposes of thermodynamics has been  
to provide a characterization of systems in terms of only a few  parameters. 
Note, however, that the thermodynamic parameters such as $\beta$ are only
given on a two--surface $B$. {\it No} assertion is made about the values 
of $\beta$ inside $B$. Thus the generalized thermodynamical description 
required is still 
{\it far} simpler than that of the full configuration of the system. The 
thermodynamical formalism that results from a generalization to 
non--constant thermodynamical data also has a number of compelling 
features. In particular, the thermodynamical data are brought into 
direct correspondence with the canonical 
boundary data, and in the process an intimate connection between
thermodynamics and dynamics is revealed [11--13]. 
%%%%%%%%%%%%%%%%%%%%%%%%%%%%%%%%%%%%%%%%%%%%%%%%%%%%%%%%%%%%%%%% 

\bigskip
\addtocounter{chapter}{1}
\setcounter{equation}{0}
\vspace{10pt}
\centerline{\bf III. The Action} 
\vspace{4pt}
Assume that the spacetime manifold~$\cal M$ is topologically the 
product of a spacelike hypersurface and a real line 
interval, $\Sigma\times I$. The boundary of $\Sigma$ is denoted 
$\partial\Sigma = B$. The spacetime metric is $g_{\mu\nu}$ with 
associated curvature tensor ${\cal R}_{\mu\nu\sigma\rho}$ and derivative 
operator $\nabla_{\mu}$. The boundary of $\cal M$, $\partial\cal M$, 
consists of initial and final spacelike hypersurfaces $t'$ and $t''$, 
respectively, and a timelike hypersurface $\Bthree = B\times I$ joining 
these. The induced metric on the spacelike hypersurfaces $t'$ and $t''$ 
is denoted by $h_{ij}$, and the induced metric on $\Bthree$ is denoted by 
$\gamma_{ij}$.\footnote{We use latin letters $i$, $j$, $k$, $\ldots$ as 
indices both for tensors on $\Bthree$ and for tensors on a generic
hypersurface 
$\Sigma$. The two uses of such indices can be distinguished by the 
context in which they occur.} 

Consider the gravitational action 
\be 
  S^1 = \frac{1}{2\kappa} \int_{\cal M}d^4x\sqrt{-g}({\cal R}-2\Lambda) 
  + \frac{1}{\kappa} \int_{t'}^{t''}d^3x \sqrt{h}\,K 
  - \frac{1}{\kappa} \int_{{}^3\! B} d^3x \sqrt{-\gamma}\,\Theta \ . 
\ee
Here, $\kappa$ is $8\pi$ times Newton's constant and $\Lambda$ is the 
cosmological constant. For simplicity, we have omitted matter 
contributions to the action. The symbol $\int_{t'}^{t''} d^3x$ denotes 
an integral over the boundary element $t''$ minus an integral over the 
boundary element $t'$. The function $K$ is the trace of the extrinsic 
curvature $K_{ij}$ for the boundary elements $t'$ and $t''$, defined 
with respect to the future pointing unit normal $u^\mu$. Likewise, 
$\Theta$ is the trace of the extrinsic curvature $\Theta_{ij}$ of the 
boundary element $\Bthree$, defined with respect to the outward pointing 
unit normal $n^\mu$. 

Under variations of the metric the action (3.1) varies according to 
\bea 
   \delta S^1 & = & \hbox{(terms that vanish when the equations of
    motion hold)}\nonumber\\ 
    & & + \int_{t'}^{t''} d^3x\,P^{ij}\delta h_{ij} +
    \int_{{}^3\! B} d^3x\,\pi^{ij}\delta \gamma_{ij} 
    - \frac{1}{\kappa} \int_{B'}^{B''} d^2x \sqrt{\sigma} \delta\alpha 
    \ . 
\eea
The coefficient of $\delta h_{ij}$ in the boundary terms at $t'$ and 
$t''$ is the gravitational momentum 
\be 
   P^{ij} = \frac{1}{2\kappa} \sqrt{h} \bigl( K h^{ij} - K^{ij} \bigr) \ . 
\ee 
Likewise, the coefficient of $\delta\gamma_{ij}$ in the boundary term 
at $\Bthree$ is 
\be 
   \pi^{ij} = -\frac{1}{2\kappa} \sqrt{-\gamma} \bigl( \Theta \gamma^{ij} 
    - \Theta^{ij} \bigr) \ . 
\ee 
Equation (3.2) also includes 
integrals over the ``corners" $B'' = t''\cap\Bthree$ and 
$B' = t'\cap\Bthree$ whose integrands are proportional to the variation of
the 
``angle" $\alpha = \sinh^{-1}(u\cdot n)$ between the unit normals $u^\mu$ of 
the hypersurfaces $t''$ and $t'$ and the unit normal $n^\mu$ of 
$\Bthree$ [14,15]. The determinant of the two--metric on $B'$ or $B''$ is 
denoted by $\sigma$. 

The action $S^1$ yields the classical equations of motion when the 
induced metric on $\Bthree$, $t'$, and $t''$ and the angle $\alpha$ at $B'$
and 
$B''$ are held fixed in the variational principle. In general, the 
functional $S=S^1-S^0$, where $S^0$ is a functional of the fixed boundary 
data, also yields the classical equations of motion. For simplicity we 
define $S^0$ to be a functional of $\gamma_{ij}$ only. The variation 
$\delta S$ then differs from $\delta S^1$ of Eq.~(3.2) only in that 
$\pi^{ij}$ is replaced by $\pi^{ij} - (\delta S^0/\delta\gamma_{ij})$. 

Now foliate the boundary element $\Bthree$ into two--dimensional 
surfaces $B$ with induced two--metrics $\sigma_{ab}$. The 
three--metric $\gamma_{ij}$ can be written according to the familiar 
Arnowitt--Deser--Misner decomposition as 
\be 
   \gamma_{ij}\,dx^idx^j=-N^2dt^2+\sigma_{ab}
   (dx^a+V^a dt)(dx^b+V^b dt) \ , 
\ee
where $N$ is the lapse function and $V^a$ is the shift vector. The 
corresponding variation of $\gamma_{ij}$ is [10] 
\be 
   \delta\gamma_{ij} = (-{2} u_i u_j /N) \delta N + (-{2} 
   \sigma_{a{\sss (}i}
  u_{j{\sss )}} /N)\delta V^a + (\sigma_{{\sss (}i}^{a} \sigma_{j{\sss
  )}}^{b})  
    \delta\sigma_{ab} \ ,  
\ee
where $u_i$ is the unit normal of the slices $B$ and 
$\sigma^i_a = \delta^i_a$ projects covariant tensors from $\Bthree$ to 
the slices $B$. With this result, the 
contribution to the variation of $S$ from the boundary element $\Bthree$ 
becomes 
\bea 
    \delta S\bigr|_{{}^3\! B} & = & \int_{{}^3\! B} d^3x \bigl(\pi^{ij} - 
   (\delta S^0/\delta\gamma_{ij}) \bigr) \delta \gamma_{ij} \nonumber \\
    & = & \int_{{}^3\! B} d^3x \sqrt{\sigma}\Bigl( -\varepsilon\delta N + 
    j_a\delta V^a + (N/2)s^{ab}\delta\sigma_{ab} \Bigr) \ ,
\eea
where the coefficients of the varied fields are defined by 
\bea 
   \varepsilon & = & \frac{2}{N\sqrt{\sigma}} u_i\pi^{ij} u_j + 
    \frac{1}{\sqrt{\sigma}} \frac{\delta S^0}{\delta N} \ ,\\
    j_a & = & -\frac{2}{N\sqrt{\sigma}} \sigma_{ai} \pi^{ij} u_j - 
    \frac{1}{\sqrt{\sigma}} \frac{\delta S^0}{\delta V^a} \ ,\\
    s^{ab} & = & \frac{2}{N\sqrt{\sigma}} \sigma_{i}^a \pi^{ij} \sigma_j^b  
    - \frac{2}{N\sqrt{\sigma}} \frac{\delta S^0}{\delta\sigma_{ab}} 
    \ . 
\eea
The leading terms in Eqs.~(3.8)--(3.10) can be rewritten in terms of an 
extrinsic curvature $k_{ab}$ that is defined by parallel transporting the 
unit normal $n^\mu$ of $\Bthree$ across a two--dimensional slice $B$. Thus, 
$k_{ab}$ is the extrinsic curvature of $B$ considered 
as the boundary $B=\partial\Sigma$ of a spacelike hypersurface $\Sigma$ 
whose unit normal $u^\mu$ is chosen orthogonal to $n^\mu$. (The case in 
which $u^\mu n_\mu\neq 0$ has been considered in Refs.~[16,17], and will 
be treated in Ref.~[15].) Also let 
$P^{ij}$ denote the gravitational momentum for the hypersurfaces 
$\Sigma$ that are ``orthogonal" to $\Bthree$, and let $a_\mu= 
u^\nu\nabla_\nu u_\mu$ denote the acceleration of the unit normal 
$u_\mu$ for this family of hypersurfaces. The resulting expressions 
are [10] 
\bea 
    \varepsilon & = & \frac{1}{\kappa} k - \varepsilon_{\sss0} 
    \ , \\
    j_i & = & -\frac{2}{\sqrt{h}}\sigma_{ij} P^{jk} n_k - (j_{\sss0})_i 
    \ , \\ 
    s^{ab} & = & \frac{1}{\kappa}\bigl( k^{ab} + 
    (n_\mu a^\mu - k)\sigma^{ab}
   \bigr)  - (s_{\sss0})^{ab} \ . 
\eea
In these equations, we have expressed the terms 
proportional to the functional derivatives of $S^0$ as $\varepsilon_{\sss
0}$, 
$(j_{\sss 0})_i$, and $(s_{\sss 0})^{ab}$. We will assume that $S^0$ is a 
linear functional of the lapse $N$ and shift $V^a$, so that 
$\varepsilon_{\sss 0}$ and $(j_{\sss 0})_i$ are functionals of the
two--metric 
$\sigma_{ab}$ only [10]. Also note that the indices in 
Eq.~(3.12) refer to the hypersurface $\Sigma$. Thus, $j_i = j_a\sigma^a_i$ 
where $\sigma^i_a = \delta^i_a$ projects tensors from $\Sigma$ to $B$, and 
$\sigma_{ij} = \sigma^a_i\sigma_{aj}$. 

From its definition through Eq.~(3.7), $\sqrt{\sigma}\varepsilon$ equals
minus the time rate of change of the action, where changes in time are 
controlled by the lapse function $N$ on $\Bthree$. Thus, $\varepsilon$ 
is identified as an energy surface density for the system [10]. Likewise, 
we  identify $j_i$ as the momentum surface density and $s^{ab}$ as the 
spatial stress [10]. The fact that the relevant stress--energy--momentum 
quantities live on the boundary of space is a consequence of the fact 
that the gravitational field on the boundary $B$ ``knows" about the 
entire matter--energy 
content of space interior to $B$. Also note that the total energy $E$ can be 
defined by integrating $\sqrt{\sigma}\varepsilon$ over the two--surface $B$. 
The result is precisely the energy discussed in Sec.~2. In particular, if 
$B$ is a spherical boundary in a static time slice of a Schwarzschild black 
hole, the total energy $E$ is given by Eq.~(2.8). (This assumes $S^0$ is
chosen such that $\varepsilon_{\sss 0} = -1/(4\pi R)$.) 
%%%%%%%%%%%%%%%%%%%%%%%%%%%%%%%%%%%%%%%%%%%%%%%%%%%%%%%%%%%%%%%%%

\bigskip
\addtocounter{chapter}{1}
\setcounter{equation}{0}
\vspace{10pt}
\centerline{\bf IV. Lorentzian and Euclidean Notation} 
\vspace{4pt} 
The action $S$ discussed in the previous section can be viewed as a
functional of {\it complex} metrics. In particular $S$ is the action 
for both Lorentzian and Euclidean metrics.  We adopt the point of 
view that, strictly speaking, there is no distinction between 
the ``Lorentzian action" and the ``Euclidean action", or between the
``Lorentzian 
equations of motion" and the ``Euclidean equations of motion". 
Of course, a particular {\it solution\/} of the classical 
equations of motion can be Lorentzian or Euclidean. But for the 
action functional itself, the only distinction between Lorentzian 
and Euclidean is simply one of 
{\it notation\/}. In Sec.~3 we have used what might be called Lorentzian 
notation: the action $S$ is defined with the convention that $\exp(iS)$ 
is the phase in the path integral; the volume 
elements for ${\cal M}$ and $\Bthree$ are written as $\sqrt{-g}$ and 
$\sqrt{-\gamma}$, respectively; the lapse function associated with the
foliation 
of ${\cal M}$ into hypersurfaces $\Sigma$ is defined by $N \equiv
\sqrt{-1/g^{tt}}$. (The 
lapse function that appears in Eq.~(3.5) is the restriction of this 
spacetime lapse to the boundary element $\Bthree$. It is defined by 
$N \equiv \sqrt{-1/\gamma^{tt}}$.)  Therefore $S$, $\sqrt{-g}$, 
$\sqrt{-\gamma}$, and $N$ are real for Lorentzian metrics and 
imaginary for Euclidean metrics. We can re--express the action in Euclidean 
{\it notation} by making the following changes. Define a new 
action functional by $I[g] \equiv -i S[g]$ so that the phase in the path
integral 
is given by $\exp(-I)$. Also rewrite the volume elements for ${\cal M}$ and 
$\Bthree$ as $\sqrt{g} \equiv i\sqrt{-g}$ and $\sqrt{\gamma} \equiv
i\sqrt{-\gamma}$, 
respectively, and define a new lapse function by $\Nbar \equiv
\sqrt{1/g^{tt}} 
\equiv i\sqrt{-1/g^{tt}} \equiv i N$.\footnote{A bit of care is required in
defining 
the square roots. For example, the appropriate definition of $\sqrt{-g}$ is 
obtained by taking the branch cut in the upper half complex plane, say, along
the positive imaginary axis. Then the imaginary part of $\sqrt{-g}$ is 
negative. Correspondingly, the appropriate definition of $\sqrt{g}$ is 
obtained by taking the 
branch cut along the negative imaginary axis. Then the imaginary part of
$\sqrt{g}$ is positive.} 

It is also 
convenient to redefine the timelike unit normal of the 
slices $\Sigma$. In Lorentzian notation, the unit normal is defined by 
$u_\mu \equiv -N\delta_\mu^t$ and satisfies $u\cdot u = -1$.  
A new unit normal is defined by $\ubar_\mu \equiv \Nbar\delta_\mu^t 
\equiv i N\delta_\mu^t \equiv -i u_\mu$, and satisfies $\ubar\cdot \ubar =
+1$. 
In some contexts it is also useful to define a new extrinsic curvature 
$\Kbar_{\mu\nu}$ in terms of the normal $\ubar_\mu$. 
$\Kbar_{\mu\nu}$ is related to the old extrinsic curvature $K_{\mu\nu}$ 
by $\Kbar_{\mu\nu} \equiv 
-(\delta^\sigma_\mu - \ubar^\sigma \ubar_\mu)\nabla_\sigma \ubar_\nu \equiv 
i (\delta^\sigma_\mu + u^\sigma u_\mu)\nabla_\sigma u_\nu \equiv  -i
K_{\mu\nu}$. 
In turn, $\Kbar_{\mu\nu}$ can be used to define a new gravitational momentum
$\Pbar^{ij}$ that is related to the momentum of Eq.~(3.3) by 
$\Pbar^{ij} \equiv -i P^{ij}$. We will, however, continue to use the old 
notations $K_{ij}$ and $P^{ij}$. 

In addition to the notational changes described above, we will also define 
a new shift vector by $\Vbar_i \equiv i g_{ti} \equiv i V_i$. This 
notation is a departure from the standard Euclidean notation in the sense
that 
$\Vbar_i$ {\it is imaginary for Euclidean 
metrics\/}. One of the motivations for this change is the following. Apart
from surface 
terms, the gravitational Hamiltonian is a linear combination of constraints 
built from the gravitational canonical data with the lapse function and 
shift vector as coefficients. In conjunction with the new notation 
$\Nbar$, $\Vbar_i$ 
for the lapse and shift, we choose to continue to denote the gravitational
canonical 
data by $h_{ij}$, $P^{ij}$, as mentioned above. Then the constraints are 
unaffected by the change in notation, and the Hamiltonian can be written as 
$H[N,V] \equiv -i H[\Nbar,\Vbar]$. The overall factor of $-i$ that appears in
this relationship is precisely what is 
required for the connection between the evolution operator ($e^{-iHt}$ in 
particle mechanics) and the density operator ($e^{-H\beta}$ in ordinary 
statistical mechanics). When the gravitational field is coupled to other 
gauge fields, such as Yang--Mills or electrodynamics, it is natural to 
redefine the Lagrange multipliers associated with the gauge constraints as 
well [11]. 

With our new notation, Eq.~(3.1) becomes 
\be 
   I^1 = -\frac{1}{2\kappa} \int_{\cal M}d^4x\sqrt{g}({\cal R}-2\Lambda) 
  - \frac{i}{\kappa} \int_{t'}^{t''}d^3x \sqrt{h}\,K  
  + \frac{1}{\kappa} \int_{{}^3\! B} d^3x \sqrt{\gamma}\,\Theta  
\ee
and Eq.~(3.2) becomes 
\bea 
   \delta I^1 & = & \hbox{(terms that vanish when the equations of
    motion hold)}\nonumber\\ 
    & & -i  \int_{t'}^{t''} d^3x\, P^{ij}\delta h_{ij} -i
    \int_{{}^3\! B} d^3x\, \pi^{ij}\delta \gamma_{ij} 
    + \frac{1}{\kappa} \int_{B'}^{B''} d^2x \sqrt{\sigma} \delta\alphabar 
    \ . \quad
\eea
Here, we have defined $\alphabar \equiv \cos^{-1}(\ubar\cdot n)$ so that 
$\delta\alphabar \equiv i\delta\alpha$. Thus, $\alphabar$ is the angle 
between the unit normals $\ubar$ and $n$ of the boundary elements $t''$ 
(or $t'$) and $\Bthree$. The full action $I\equiv -iS$ differs from $I^1$ 
by a term $I^0\equiv -iS^0$ that is a functional of the metric $\gamma_{ij}$ 
on $\Bthree$. The contribution to $\delta I$ from the boundary element 
$\Bthree$ is obtained from Eq.~(3.7). Putting this together with Eq.~(4.2) 
yields 
\bea 
   \delta I & = & \hbox{(terms that vanish when the equations of motion 
       hold)}\nonumber\\ 
    & & -i  \int_{t'}^{t''} d^3x\, P^{ij}\delta h_{ij} + \frac{1}{\kappa} 
      \int_{B'}^{B''} d^2x \sqrt{\sigma} \delta\alphabar \nonumber\\ 
    & & +\int_{{}^3\! B} d^3x \sqrt{\sigma} \left( \varepsilon\delta\Nbar 
      - j_a\delta\Vbar^a - (\Nbar/2) s^{ab}\delta\sigma_{ab} \right) 
    \ . \quad
\eea
The stress--energy--momentum quantities $\varepsilon$, $j_i$, and $s^{ab}$ 
from Eqs.~(3.11)--(3.13) are given by 
\bea 
    \varepsilon & = & \frac{1}{\kappa} k - \frac{1}{\sqrt{\sigma}} 
    \frac{\delta I^0}{\delta\Nbar} \ , \\
    j_i & = & -\frac{2}{\sqrt{h}}\sigma_{ij} P^{jk} n_k + 
    \frac{\sigma_i^a}{\sqrt{\sigma}} \frac{\delta I^0}{\delta\Vbar^a} \ , \\ 
    s^{ab} & = & \frac{1}{\kappa} ( k^{ab}  - k\sigma^{ab}) + \left(
    \frac{n^i\partial_i \Nbar}{\kappa\Nbar}\right) \sigma^{ab} + 
    \frac{2}{\Nbar\sqrt{\sigma}} \frac{\delta I^0}{\delta\sigma_{ab}} \ . 
\eea
In Eq.~(4.6) we have written the $n_\mu$ component of the 
acceleration of the unit normal $u_\mu$ as 
$n_\mu a^\mu \equiv (n^i\partial_i N)/N \equiv 
(n^i\partial_i \Nbar)/\Nbar$. Also note that the 
action $I$ in canonical form is [10,12] 
\be 
   I = \int_{\cal M} d^4x \left( -i P^{ij} {\dot h}_{ij} + \Nbar{\cal H} 
   + \Vbar^i{\cal H}_i \right) + \int_{{}^3\! B} d^3x \sqrt{\sigma} \left( 
   \Nbar\varepsilon - \Vbar^a j_a \right) \ .
\ee 
Here, ${\cal H}$ and ${\cal H}_i$ are the Hamiltonian and momentum
constraints.

Since $I^0$ (like $S^0$) is functionally linear in the lapse $\Nbar$ and
shift $\Vbar^a$, then $\varepsilon$ and $j_i$ are functions of the 
canonical variables only. 
Moreover, $s^{ab}$ depends on the lapse and shift only through ratios 
such as $(\partial_i \Nbar)/\Nbar$ and $\Vbar^a/\Nbar$. These 
observations are important 
for the following reason. Consider the metric $g_{\sss L}$ for a stationary 
Lorentzian geometry, written in stationary coordinates. Given such a metric,
we can compute the stress--energy--momentum quantities $\varepsilon$, 
$j_i$, and $s^{ab}$ for the surface $\Bthree$ that is 
defined by taking a two--surface $B$ within one of the stationary time 
slices and extending this surface along the stationary time direction. 
Now consider the 
generally complex metric $g_{\sss C}$ obtained by substituting $t\to -it$ 
in the stationary Lorentzian metric $g_{\sss L}$. Under this 
transformation the lapse 
function $N$ and shift vector $V^i$ are changed from real to imaginary, and
the canonical variables $h_{ij}$, $P^{ij}$ are invariant. That is, 
the substitution $t\to -it$ generates a complex metric $g_{\sss C}$ with 
$\Nbar$, $\Vbar^i$, $h_{ij}$, and $P^{ij}$ real. (The metric will be 
Euclidean if $V^i=0$.) If we now compute $\varepsilon$, 
$j_i$, and $s^{ab}$ for the surface $\Bthree$ for the complex metric, 
the results will be the same as the results for the original Lorentzian
metric. 
This observation is the key to understanding the relationship between the
results 
of a functional integral over complex (or Euclidean) metrics and the real 
Lorentzian spacetime that it describes [12]. For example, 
suppose we compute the partition function as a functional integral and find
that 
it is extremized by a certain stationary complex (or Euclidean) black hole
metric $g_{\sss C}$.\footnote{The boundary data for the partition function 
is chosen to be stationary on $\Bthree$, so the extremum of the action 
can be expected to be stationary as well.} The partition function 
characterizes the system in terms 
of its thermal properties, such as its temperature, its expectation value 
for energy, {\it etc\/}. Can we conclude that the 
physical system described by this partition function is a real Lorentzian
black hole simply because the path integral is extremized by a complex 
(or Euclidean) black hole? The answer is yes, precisely because the 
stress--energy--momentum for the stationary complex black hole 
$g_{\sss C}$ coincides with the 
stress--energy--momentum of the related stationary Lorentzian black hole
$g_{\sss L}$. 
Thus, for example, when we compute the expectation value of energy from the 
partition function, it will coincide in the zero--loop approximation with
the energy surface density $\varepsilon$ of the complex black hole 
$g_{\sss C}$, which in turn characterizes the energy surface density of a 
real Lorentzian black hole 
$g_{\sss L}$.  In this way we can conclude that the partition function 
indeed provides a description of the thermal properties of a physical black
hole. 
%%%%%%%%%%%%%%%%%%%%%%%%%%%%%%%%%%%%%%%%%%%%%%%%%%%%%%%%%%%%%%%%%%%

\bigskip
\addtocounter{chapter}{1}
\setcounter{equation}{0}
\vspace{10pt}
\centerline{\bf V. Functional Integrals} 
\vspace{4pt}
A path integral constructed from an action $I$ is a functional of the
quantities 
that are held fixed in the variational principle $\delta I = 0$. What are
held 
fixed in the variational principle are the quantities that appear varied in
the 
boundary terms of $\delta I$. The fixed boundary data for the action of 
Eq.~(4.3) are the metric $h_{ij}$ on $t'$ and $t''$, the angle 
$\alphabar$ at the corners $B'$ and $B''$, and the lapse function $\Nbar$,
shift vector $\Vbar^a$, and two--metric $\sigma_{ab}$ on $\Bthree$. 
In the path integral, 
the gauge invariant part of the data on $\Bthree$ corresponds to the inverse 
temperature $\beta$, chemical potential $\omega^a$, and two--geometry of the 
boundary $B$. These are grand canonical boundary conditions. 

The inverse temperature is defined in terms of the boundary data on 
$\Bthree$ by [12] 
\be 
   \beta = \int dt\, \left. \Nbar\right|_{B} \ .
\ee 
In geometrical terms, this is the proper distance between $t'$ and $t''$ as
measured 
along the curves in $\Bthree$ that are orthogonal to the 
slices $B$. Thus, $\beta$ is the $t$--coordinate invariant part of the lapse
function 
$\Nbar$. Note that $\beta$ is a function on the space 
boundary $B$, as anticipated in the discussion at the end of Sec.~2. 

The chemical potential is defined in terms of the boundary data on $\Bthree$ 
by [12] 
\be 
   \omega^a = \frac{\int dt\, \left. \Vbar^a\right|_{B} }{\int dt\, \left. 
   \Nbar\right|_{B} } = \frac{\int dt\, \left. V^a\right|_{B} }{\int dt\,
\left. 
   N\right|_{B} } \ . 
\ee 
For a complex metric $g_{\sss C}$, $\beta\omega^a$ has a geometrical
interpretation 
as the proper distance along the $a$--coordinate line by which the spatial 
coordinates are shifted between $t'$ and $t''$. For a Lorentzian metric
$g_{\sss L}$, 
$\omega^a$ has the following interpretation. 
Recall that the shift vector $V^a$ on $\Bthree$ is the velocity 
of the spatial coordinate system with respect to the observers who are at
rest in the constant time slices $B$. It is convenient to tie the 
spatial coordinates to the local motion of the physical system [11,12], 
so that $V^a$ gives the local 
velocity of the system in terms of coordinate time. Thus, we see that the 
chemical potential $\omega^a$ is the proper velocity of the physical 
system as measured with respect to observers who are at rest at the system 
boundary $B$. In the case of axisymmetric boundary data, the only nonzero 
component of $\omega^a$ is the proper angular velocity of the system. 

The functional integral constructed from the action $I$ is 
\be 
   \rho[h'',h'; \alphabar'',\alphabar'; \beta,\omega,\sigma] = \int Dg\, 
   e^{-I[g]}  \ , 
\ee 
where $h''$ and $h'$ denote the metrics on $t''$ and $t'$, and $\alphabar''$ 
and $\alphabar'$ denote the angles at the corners $B''$ and $B'$. We will 
consider this expression as a functional integral over the 
class of complex metrics for which $\Nbar$, $\Vbar^i$, and $h_{ij}$ are 
real. 

The path integral (5.3) is the grand canonical density matrix for the
gravitational 
field in a box $B$. The grand canonical partition function, denoted 
$Z[\beta,\omega,\sigma]$, 
is obtained by tracing over the initial and final configurations. 
In the path integral language, 
this amounts to performing a periodic identification so that the manifold 
topology becomes ${\cal M} = \Sigma\times S^1$. In addition, 
$\alphabar''$ and $\alphabar'$ should be chosen so that the total angle 
$\alphabar'' + \alphabar'$ equals $\pi$. This insures that the boundary 
$\partial{\cal M}$ is smooth when the 
initial and final hypersurfaces are joined together. Thus, the 
grand canonical partition function can be written as 
\be 
   Z[\beta,\omega,\sigma] = \left. \int Dh \, \rho[h,h;
\alphabar'',\alphabar'; 
   \beta,\omega,\sigma]\right|_{\alphabar'' + \alphabar' = \pi} \ . 
\ee 
The right hand side of this expression apparently depends on the 
angle difference $\alphabar'' - \alphabar'$. However, we expect that 
with the periodic identification, $\alphabar'' - \alphabar'$ is pure 
gauge and in a more detailed analysis would be absent from the path 
integral. 

One can consider various density matrices and partition 
functions corresponding to different combinations of thermodynamical
variables, where one variable is selected from each of the conjugate 
pairs. For example, 
in ordinary statistical mechanics the thermodynamically conjugate variables 
might consist of the inverse temperature $\beta$ and energy $E$, and the 
chemical potential $\mu$ and particle number $N$. Then the grand canonical 
partition function is $Z(\beta,\mu)$, the canonical partition function is 
$Z(\beta,N)$, and the microcanonical partition function (the density of 
states) is $\nu(E,N)$. These partition functions are related to one another 
by Laplace and inverse Laplace transforms, where each transform has the
effect of switching the functional dependence from some 
thermodyanamical variable (such 
as $\beta$) to its conjugate (such as $E$). 

When the gravitational 
field is included in the description of the system, all of the
thermodynamical 
data can be expressed as boundary data [18,12]. In the path integral 
formalism, the effect of the Laplace and inverse 
Laplace transforms is simply to add or subtract certain boundary terms from 
the action. Thus, for example, the action $I_m$ appropriate for
microcanonical 
boundary conditions just differs from the action $I$ by boundary terms: 
\be 
   I_m = I - \int_{{}^3\! B} d^3x \sqrt{\sigma} \left( \Nbar \varepsilon
      - \Vbar^a j_a \right) \ . 
\ee 
From Eq.~(4.3) we see that the contribution to $\delta I_m$ from the 
boundary element $\Bthree$ is 
\be 
   \left. \delta I_m \right|_{{}^3\! B} = \int_{{}^3\! B} d^3x \left( 
   - \Nbar\delta(\sqrt{\sigma}\varepsilon) + \Vbar^a\delta(\sqrt{\sigma} 
   j_a) - (\Nbar\sqrt{\sigma}/2) s^{ab} \delta\sigma_{ab} \right) \ . 
\ee 
The path integral constructed from $I_m$ is a functional of the quantities 
that appear varied in the boundary terms of $\delta I_m$. Thus, the 
microcanonical density matrix is 
\be 
   \rho_m[h'',h'; \alphabar'',\alphabar'; \varepsilon, 
    j,\sigma] = \int Dg\,  e^{-I_m[g]}  \ , 
\ee 
The trace of this density matrix is the density of states for the
gravitational 
field in a box, 
\be 
   \nu[ \varepsilon, j,\sigma] = \left.\int Dh \, \rho_m[h,h;
\alphabar'',\alphabar'; 
   \varepsilon, j,\sigma] \right|_{\alphabar'' + \alphabar' = \pi} \ . 
\ee 
When combined, Eqs.~(5.7) and (5.8) yield a path integral expression for 
the density of states. If the contours of integration for 
$\Nbar$ and $\Vbar^i$ are rotated from the real to the imaginary axis, then 
$\nu[ \varepsilon, j,\sigma]$ is expressed as a functional integral over 
the class of metrics with $N$, $V^i$, and $h_{ij}$ real. In other 
words, $\nu[ \varepsilon, j,\sigma]$ can be expressed as a path integral 
over real Lorentzian metrics [12]. 
%%%%%%%%%%%%%%%%%%%%%%%%%%%%%%%%%%%%%%%%%%%%%%%%%%%%%%%%%%%%%%% 

\bigskip
\addtocounter{chapter}{1}
\setcounter{equation}{0}
\vspace{10pt}
\centerline{\bf VI. Black Hole Entropy} 
\vspace{4pt} 
In the previous section the partition function $Z[\beta,\omega,\sigma]$ and 
the density of states $\nu[\varepsilon, j,\sigma]$ were constructed as
functional 
integrals over the gravitational field on manifolds whose topologies are  
necessarily $\Sigma\times S^1$. This would seem to be an unavoidable
consequence 
of deriving $Z[\beta,\omega,\sigma]$ and $\nu[\varepsilon, j,\sigma]$ from 
traces of density matrices because the density matrices $\rho$ and $\rho_m$ 
are defined in terms of functional integrals on manifolds ${\cal M}$ 
with product topology $\Sigma\times I$. However, experience has shown that
for a black hole, the functional integrals for the partition function and 
density of states are extremized by a complex (or Euclidean) metric on 
the manifold $R^2\times S^2$ [3,12]. 
Thus, one  would expect the black hole contribution to the density of states 
to come from a path integral that is defined on a manifold with topology 
$R^2\times S^2$. In this section we will show how the black hole density of 
states can be related to the microcanonical density matrix. 

We begin by considering the manifold ${\cal M} = \Sigma\times I$, where
$\Sigma$ 
is topologically a thick spherical shell ($S^2\times I$). The boundary 
$\partial\Sigma = B$ 
consists of two disconnected surfaces, an inner sphere $B_i$ and an outer 
sphere $B_o$. The boundary element $\Bthree$ consists of disconnected 
surfaces as well, $\Bthree_i = B_i\times I$ and $\Bthree_o = B_o\times I$. 
The results of the previous sections can be applied in constructing various 
density matrices for the gravitational field on $\Sigma$. We wish to 
consider the particular density matrix $\rho_*$ that is defined through 
the path integral by the action 
\be 
   I_* = I - \int_{{}^3\! B_o} d^3x \sqrt{\sigma} \left( \Nbar \varepsilon
      - \Vbar^a j_a \right) + \int_{{}^3\! B_i} d^3x \sqrt{\sigma} \Nbar 
      s^a_a/2 \ . 
\ee 
$I_*$ differs from the action $I$ of Sec.~4 by boundary terms that are not
the 
same for the two disconnected parts of $\Bthree$. The contributions to the 
variation $\delta I_*$ from $\Bthree_i$ and $\Bthree_o$ are 
\bea 
   \left. \delta I_* \right|_{{}^3\! B_i} & = & \int_{{}^3\! B_i} d^3x 
   \left(
  (\sqrt{\sigma}\varepsilon)\delta\Nbar - (\sqrt{\sigma}
j_a)\delta\Vbar^a  
      +\sigma_{ab}\delta(\Nbar\sqrt{\sigma}s^{ab}/2)  \right) \ ,\\ 
   \left. \delta I_* \right|_{{}^3\! B_o} & = & \int_{{}^3\! B_o} d^3x 
   \left(
 - \Nbar \delta(\sqrt{\sigma}\varepsilon) + \Vbar^a\delta(\sqrt{\sigma}
j_a)  
     - (\Nbar\sqrt{\sigma}s^{ab}/2)\delta\sigma_{ab} \right) \ . \qquad
\eea 
Comparison with Eq.~(5.6) shows that at the outer boundary element
$\Bthree_o$ 
we have chosen microcanonical boundary conditions. At the inner boundary 
element $\Bthree_i$ we have chosen completely open boundary conditions. 
By ``completely open" we mean that none of the traditional ``conserved" 
quantities 
like energy, angular momentum, or area is fixed. Thus, all of the 
stress--energy--momentum quantities $\varepsilon$, $j_i$, and $s^{ab}$ are 
allowed to fluctuate on the inner boundary element while their conjugates, 
the inverse temperature, chemical potential, and spatial stress, are held 
fixed. The microcanonical boundary conditions and the completely open
boundary conditions are precisely opposite in this respect. 

Observe that the spatial stress tensor $s^{ab}$ can be split into its trace 
$\theta = s^{ab}\sigma_{ab}\equiv s^a_a$, which is the sum of the normal 
components of stress (``pressures"), and the shear stresses 
$\eta^{ab} = s^{ab} - \theta\sigma^{ab}/2$. Then the contribution from 
$\Bthree_i$ to $\delta I_*$ becomes 
\bea 
    \left.\delta I_* \right|_{{}^3\! B_i} & = & \int_{{}^3\! B_i} d^3x 
    \left(
 (\sqrt{\sigma}\varepsilon)\delta\Nbar - (\sqrt{\sigma}
j_a)\delta\Vbar^a \right.
      \nonumber\\ 
      & & \qquad\qquad \left. -\sigma_{ab}\delta(\Nbar\sqrt{\sigma}
\eta^{ab}) - 
      \sqrt{\sigma}\delta (\Nbar\theta)  \right) \ . 
\eea 
Thus, the fixed data on $\Bthree_i$ consist of $\Nbar$, $\Vbar^a$, 
$\Nbar\sqrt{\sigma} \eta^{ab}$, and $\Nbar\theta$, and the  
density matrix constructed as a path integral from $I_*$ is 
\be 
   \rho_*[h'',h';\alphabar'',\alphabar'; \varepsilon,j,\sigma; 
   \Nbar,\Vbar,\Nbar\sqrt{\sigma}\eta,\Nbar\theta]  =\int Dg \, e^{-I_*[g]} 
   \ . 
\ee 
Here, it is understood that the data $\varepsilon$, $j_a$, and $\sigma_{ab}$ 
are fixed on the outer boundary element $\Bthree_o$ and the data $\Nbar$,
$\Vbar^a$, 
$\sqrt{\sigma}\eta^{ab}$, and $\theta$ are fixed on the inner boundary
element 
$\Bthree_i$. Also, $\alphabar''$ represents the angles at both disconnected
parts 
of $B''$ and likewise for $\alphabar'$. 

The black hole density of states $\nu_*[\varepsilon,j,\sigma]$ is obtained
from 
the trace of the density matrix 
$\rho_*$ along with the following special choice of data on the inner
boundary 
element $\Bthree_i$: 
\bea 
   \Nbar & = & 0 \ ,\\ 
   \Vbar^a & = & 0 \ ,\\ 
   \Nbar\sqrt{\sigma}\eta^{ab} & = & 0 \ ,\\ 
   \Nbar\theta & = & -\frac{4\pi}{\kappa(t''-t')} \ . 
\eea 
In Eq.~(6.9), $(t''-t')$ is just the range of coordinate time $t$. 
In addition to the above conditions, the sum of angles
$\alphabar''+\alphabar'$ 
must equal $\pi$ so that the boundary is smooth. The resulting expression for
the 
black hole density of states is 
\bea 
   & & \nu_*[\varepsilon,j,\sigma] \nonumber\\
   & & \qquad  =  \left.\int dh\, \rho_*[h,h;\alphabar'',\alphabar'; 
   \varepsilon,j,\sigma; 0,0,0,-4\pi/\kappa(t''-t')]
\right|_{\alphabar''+\alphabar' 
   = \pi} \ . \qquad\quad
\eea 
First we will discuss the geometrical meaning of the data
(6.6)--(6.9) and show that Eq.~(6.10) indeed gives the correct 
expression for black hole entropy. We 
will then discuss the physical implications of this result. 

With the periodic identification, condition (6.6) fixes to zero the length of
the circles on the inner boundary $\Bthree_i = B_i\times S^1$ that are 
orthogonal to the 
slices $B_i$. That is, $\Nbar = 0$ on the inner boundary causes the hole in
the planes 
defined by the unit normals $u_\mu$ and $n_\mu$ to become closed. This 
condition therefore seals the opening in the manifold, changing the manifold
topology 
to $R^2\times S^2$. The sealed inner boundary of the manifold is called the
bolt [19]. 
The condition (6.7) is not mandatory, but is a convenient choice. 
In terms of a physical black hole metric, this condition  fixes the spatial 
coordinate system to be co--rotating with the black hole [11,12]. The 
chemical potential is then related to the shift vector on the outer boundary
element 
$B_o$ according to the relationship (5.2). Without the condition (6.7), the
chemical 
potential and the shift vector on $B_o$ are not related in the simple way
(5.2). 
Condition (6.8) is chosen to be consistent with condition (6.6). Finally,
condition 
(6.9) ensures that the geometry is smooth at the bolt. This 
can be seen by writing Eq.~(6.9) in detail. From Eq.~(4.6) we have  
\be 
   -\frac{\Nbar}{\kappa}  k + \frac{2}{\kappa}  (n^i\partial_i \Nbar)   + 
    \frac{2}{\sqrt{\sigma}} \sigma_{ab}\frac{\delta I^0}{\delta\sigma_{ab}} 
    = -\frac{4\pi}{\kappa(t''-t')} \ , 
\ee 
which, in light of conditions (6.6) and (6.7), becomes 
$(t''-t') (n^i\partial_i \Nbar) = -2\pi$. Now, $(t''-t') (n^i\partial_i
\Nbar)$ is just minus the rate of change of proper circumference with 
respect to proper radius for the circles on the inner boundary 
$\Bthree_i$ that are orthogonal to the slices $B_i$. (The 
minus sign appears because $n^i$ is the unit normal to $B_i$ that points
outward from 
$\Sigma$.) Thus, condition (6.9) fixes the ratio of circumference to radius
to $2\pi$. 
This insures that the spacetime four--geometry is smooth as the inner
boundary is sealed. 

The correctness of the prescription (6.5), (6.10) can be confirmed by
considering 
the evaluation of the functional integral for 
$\nu_*[\varepsilon,j,\sigma]$, where the data $\varepsilon$, $j_a$,
$\sigma_{ab}$ on the outer boundary $\Bthree_o$ correspond to a 
stationary black hole. That is, 
let $\varepsilon$, $j_a$, $\sigma_{ab}$ be the stress--energy--momentum for 
a topologically spherical two--surface $B_o$ within a time slice of a
stationary 
Lorentzian black hole solution $g_{\sss L}$ of the Einstein equations. 
In the path integral for $\nu_*[\varepsilon,j,\sigma]$, fix this data on 
each slice $B_o$ of the outer boundary $\Bthree_o$. The path integral can 
be evaluated semiclassically by searching for metrics that extremize 
the action $I_*$ and satisfy the conditions at both $\Bthree_o$ and 
$\Bthree_i$. One such metric will be the complex metric $g_{\sss C}$ 
that is obtained by substituting $t\to -it$ in the Lorentzian black hole
solution 
$g_{\sss L}$. This complex metric will 
satisfy the boundary conditions at $\Bthree_o$ because, as discussed at the
end 
of Sec.~4, the substitution $t\to  -it$ does not affect the
stress--energy--momentum 
quantities $\varepsilon$, $j_a$, and $\sigma_{ab}$. The complex metric also
will 
satisfy the conditions at $\Bthree_i$, where $B_i$ coincides with the 
intersection of the stationary time slices and the black hole horizon for 
the Lorentzian metric. This follows from the observation that the lapse 
function corresponding to the natural stationary time--slicing vanishes 
on the horizon of a stationary black hole. Likewise, the corresponding 
shift vector vanishes on the horizon for 
co--rotating spatial coordinates. Also, with an appropriate choice of period 
$t''-t'$ for coordinate time, the complex metric will describe a smooth
geometry 
at the bolt. (A single choice of period suffices for all points on $B_i$,
because 
$n^i\partial_i \Nbar$ is proportional to the surface gravity of the
stationary 
black hole and is therefore constant across $B_i$.) 

In the zero--loop approximation to the path integral, the density of states 
becomes 
\be 
   \nu_*[\varepsilon,j,\sigma] \approx e^{-I_*[g_{\sss C}]} \ . 
\ee 
From Eqs.~(4.7) and (6.1), we see that the action $I_*$ written in canonical 
form is 
\bea 
   I_* & = & \int_{\cal M} d^4x \left( -iP^{ij} {\dot h}_{ij}  + \Nbar{\cal
H} + 
   \Vbar^i{\cal H}_i \right) \nonumber\\
   & &  + \int_{{}^3\! B_i} d^3x \sqrt{\sigma}\left( 
   \Nbar\varepsilon - \Vbar^a j_a + \Nbar\theta/2 \right) \ . 
\eea 
In evaluating this action at the solution $g_{\sss C}$, the term 
proportional to ${\dot h}_{ij}$ 
vanishes because of stationarity and the constraints ${\cal H}$ and ${\cal
H}_i$ vanish because the equations of motion are satisfied. The 
remaining terms in $I_*$ 
can  be found from the boundary conditions (6.6)--(6.9), which yield  
\be 
   I_*[g_{\sss C}] = -\frac{2\pi}{\kappa} \int_{B_i} d^2x \sqrt{\sigma} \ . 
\ee 
The integral that remains is just the area of the bolt, or the area $A_{\sss
H}$ 
of the black hole event horizon. Thus, in the approximation (6.12), the 
entropy is\footnote{The corresponding result for an arbitrary diffeomorphism 
invariant theory of gravitational and matter fields has been derived 
in Ref.~[20].} 
\be 
   {\cal S}[\varepsilon,j,\sigma] = \ln \nu_*[\varepsilon,j,\sigma] \approx 
   \frac{2\pi}{\kappa} A_{\sss H} \ .
\ee
With $\kappa = 8\pi$, this is the standard result ${\cal S} = A_{\sss H}/4$
for the 
black hole entropy. 

It might appear that the conditions (6.6)--(6.9) have been artificially 
contrived in an effort to obtain the result (6.15). On the contrary, these 
conditions are ``no boundary" conditions in the 
sense that they are precisely the conditions needed to seal the opening
$\Bthree_i$ 
in the manifold ${\cal M} =\Sigma\times S^1$, and convert the manifold
topology 
to $R^2\times S^2$. Thus, with these conditions there is no inner 
boundary in the spacetime four--geometry. Equations (6.6)--(6.9) are 
more properly called regularity conditions, rather than boundary 
conditions. Accordingly, the path integral defined through Eqs.~(6.5) 
and (6.10) is {\it not} to be viewed as a functional of any inner 
boundary data. (We took a slightly different viewpoint in Refs.~[11,12] 
which will be discussed in detail in [15].) This is why the 
argument of the density of states (6.10) 
includes only the conditions on the outer boundary $\Bthree_o$. 

For a general choice of data on $\Bthree_i$, the object defined 
by the trace of the density matrix (6.5) would have to be treated 
as a hybrid partition function with completely open boundary conditions 
on $\Bthree_i$ and microcanonical boundary conditions on $\Bthree_o$. 
That is, for a general choice of data $\Nbar$, $\Vbar^a$, 
$\sqrt{\sigma}\eta^{ab}$, $\theta$ on $\Bthree_i$, 
the trace of the density matrix $\rho_*$ yields  
\bea 
   & & Z_*[\varepsilon,j,\sigma;\Nbar, \Vbar^a, \sqrt{\sigma}\eta^{ab},
\theta] 
   \nonumber\\ 
   & & \qquad\quad  = 
   \left.\int dh\, \rho_*[h,h;\alphabar'',\alphabar'; 
   \varepsilon,j,\sigma; \Nbar, \Vbar^a, \sqrt{\sigma}\eta^{ab}, \theta] 
   \right|_{\alphabar''+\alphabar' = \pi} \ . \qquad\qquad
\eea 
In order to compute the density of states in this case, we would first
perform a series of inverse Laplace transforms to change the data on 
$\Bthree_i$ from completely open conditions to microcanonical conditions. 
The inverse Laplace transforms 
would have the effect of canceling the boundary terms in the action (6.13) 
and changing the action to the microcanonical action $I_m$ of Eq.~(5.5). 
This action is zero for any stationary solution, including stationary 
black holes. So in the zero--loop approximation, the entropy so computed 
would be zero. This holds true even if the action is extremized by a 
black hole solution. 

The difference between the density of states as derived from the 
partition function $Z_*$ and the black hole density of states $\nu_*$ is
precisely 
in the presence or absence of an inner boundary $\Bthree_i$ and its
associated 
data. Thus, one is led to the view that the black hole entropy derived from 
$\nu_*$ is associated with a {\it lack} or absence of boundary
information. 
The entropy derived 
from $\nu_*$ is greater than the entropy derived from $Z_*$ because in the 
case of $\nu_*$ there is less spacetime boundary data information: less 
information, more entropy. 

We should remark that our present interpretation does not support 
the view [5] that the entropy derived from $\nu_*$ arises from an 
integration over interior boundary data. In fact, by integrating 
over interior data 
in the density of states derived from $Z_*$, one merely recovers the
partition 
function $Z_*$. The black hole density of states $\nu_*$ should be viewed 
as the result of a path integral over certain gravitational fields
(spacetimes) 
with no inner boundaries whatsoever, not as the result of an integration 
over inner boundary conditions. 

The above conclusions will be developed in greater detail in 
forthcoming publications [15]. 

%%%%%%%%%%%%%%%%%%%%%%%%%%%%%%%%%%%%%%%%%%%%%%%%%%%%%%%%%%%%%%
\bigskip
\vspace{5pt}  
\centerline{\bf Acknowledgments} 
\vskip3pt 
We would like to thank S. Carlip and C. Teitelboim for stimulating
discussions. 
This research was partially supported by the National Science Foundation 
grants PHY--8908741 and PHY--9413207. 

%%%%%%%%%%%%%%%%%%%%%%%%%%%%%%%%%%%%%%%%%%%%%%%%%%%%%%%%%%%%%%%%%%
\bigskip
\vspace{.1in} 
\centerline{\bf References}  
\vspace{-.1in} 
\begin{enumerate} 
%%1:
\item J.D. Bekenstein, {\em Phys. Rev.} {\bf D7}, 2333 (1973). 
%%2: 
\item S.W. Hawking, {\em Commun. Math. Phys.} {\bf 43}, 199 (1975). 
%%3:
\item G.W. Gibbons and S.W. Hawking, {\em Phys. Rev.} {\bf D15}, 2752 
(1977); S.W. Hawking in {\em General Relativity}, edited by S.W. 
Hawking and W. Israel (Cambridge University Press, Cambridge, 1979). 
%%4: 
\item J.D. Brown, J. Creighton, and R.B. Mann, {\em Phys. Rev.} 
{\bf D50}, 6394 (1994). 
%%5: 
\item S. Carlip and C. Teitelboim, {\em Class. Quantum Grav.} {\bf 12}, 
1699 (1995). See also M. Ba\~{n}ados, C. Teitelboim, and J. Zanelli, {\em 
Phys. Rev. Lett.} {\bf 72}, 957 (1994). 
%%6:
\item See for example H.B. Callen, {\em Thermodynamics and an Introduction 
to Thermostatistics} (John Wiley \& Sons, New York, 1985). 
%%7:  
\item D.J. Gross, M.J. Perry, and L.G. Yaffe, {\em Phys. Rev.} {\bf D25}, 
330 (1982). 
%%8: 
\item J.W. York, {\em Phys. Rev.} {\bf D33}, 2029 (1986). 
%%9: 
\item R.C. Tolman, {\em Phys. Rev.} {\bf 35}, 904 (1930). 
%%10: 
\item J.D. Brown and J.W. York, {\em Phys. Rev.} {\bf D47}, 1407 (1993). 
%%11: 
\item J.D. Brown, E.A. Martinez, and J.W. York, {\em Phys. Rev. Lett.} 
{\bf 66}, 
2281 (1991); J.D. Brown and J.W. York, in {\em Physical Origins of Time
Asymmetry\/}, 
edited by J.J. Halliwell, J. Perez--Mercader, and W. Zurek (Cambridge
University Press, Cambridge, 1994) 465. 
%%12: 
\item J.D. Brown and J.W. York, {\em Phys. Rev.} {\bf D47}, 1420 (1993). 
%%13: 
\item B.F. Whiting and J.W. York, {\em Phys. Rev. Lett.} {\bf 61}, 1336
(1988). 
%%14:
\item G. Hayward, {\em Phys. Rev.} {\bf D47}, 3275 (1993). 
%%15: 
\item J.D. Brown, S.R. Lau, and J.W. York, manuscripts in preparation.  
%%16: 
\item G. Hayward and K. Wong, {\em Phys. Rev.} {\bf D46}, 620 (1992); 
{\em Phys. Rev.} {\bf D47}, 4778 (1993). 
%%17: 
\item S. Lau, ``Quasilocal Energy and Canonical Variables in General 
Relativity" (UNC Ph.D. thesis, April 1994, unpublished). 
%%18: 
\item J.D. Brown, G.L. Comer, E.A. Martinez, J. Melmed, B.F. Whiting, and
J.W. York, 
{\em Class. Quantum Grav.} {\bf 7}, 1433 (1990). 
%%19: 
\item G.W. Gibbons and S.W. Hawking, {\em Commun. Math. Phys.} {\bf 66}, 291
(1979). 
%%20: 
\item J.D. Brown, to be published in {\em Phys. Rev.} {\bf D} 
(gr-qc/9506085). 
\end{enumerate}
%%%%%%%%%%%%%%%%%%%%%%%%%%%%%%%%%%%%%%%%%%%%%%%%%%%%%%%%%%%%%%%%%%
\end{document}